\documentclass[11pt,a4paper,final]{iopart}
\usepackage{iopams}
\usepackage{graphicx}
\usepackage{cite,xcolor}
\usepackage{scalerel}
\usepackage{stackengine,wasysym}

\newcommand\reallywidetilde[1]{\ThisStyle{%
  \setbox0=\hbox{$\SavedStyle#1$}%
  \stackengine{-.1\LMpt}{$\SavedStyle#1$}{%
    \stretchto{\scaleto{\SavedStyle\mkern.2mu\AC}{.5150\wd0}}{.6\ht0}%
  }{O}{c}{F}{T}{S}%
}}

\def\hugetilde#1{$%
  \reallywidetilde{#1}\,
$\par}

\begin{document}

\title[A unified time scale for quantum chaotic regimes]
      {A unified time scale for quantum chaotic regimes}

\author{Ignacio S. Gomez$^{1}$}
\ead{nachosky@fisica.unlp.edu.ar}

\author{Ernesto P. Borges$^{1}$}
\ead{ernesto@ufba.br}

\address{$ $ \\ $^1$Instituto de F\'{i}sica, Universidade Federal da Bahia,
         Rua Barao de Jeremoabo, 40170-115 Salvador--BA, Brazil}

\begin{abstract}
We present a generalised time scale
for quantum chaos dynamics,
motivated by nonextensive statistical mechanics.
It recovers, as particular cases, the relaxation (Heisenberg)
and the random (Ehrenfest) time scales.
Moreover, we show that the generalised time scale can also be obtained
from a nonextensive version of the Kolmogorov-Sinai entropy
by considering the graininess of quantum phase space
and a generalised uncorrelation between subsets of the phase space.
Lyapunov and regular regimes for the fidelity decay are obtained
as a consequence of a nonextensive generalisation of the $m$th point
correlation function for a uniformly distributed perturbation
in the classical limit.
\end{abstract}

\vspace{2pc}
\noindent{\it Keywords}: quantum chaos,
time scales,
Kolmogorov-Sinai entropy,
fidelity.

\section{\label{sec:intro}Introduction}

The characteristic time scales are important indicators
for describing the dynamics of quantum chaotic systems.
They allow to distinguish the regular behaviour
(relaxation, or Heisenberg, time scale)
from the chaotic behaviour
(random, or logarithmic, or Ehrenfest, time scale)
in such a way as to make compatible with the Correspondence Principle (CP)
with the discrete spectrum.
One of their main features is that the quantum and classical descriptions
tend to coincide within these time scales, making possible to characterise
the phenomena of relaxation, exponential sensitivity, etc
\cite{Ber89,Cas95,Gut90,Haa01,Sto99}.
The random time scale establishes the time interval for which the dynamics
of a wavepacket is as random as the classical trajectory,
exhibiting a spreading over the whole phase space \cite{Cas95}.
The relaxation time scale, on the other hand,
establishes the minimum time interval for determining a discrete spectrum.
Some authors consider that the random time scale solves the apparent conflict
between the CP and the quantum to classical transition
in chaotic dynamics \cite{Cas95}.

Other peculiarities of quantum systems with a chaotic behaviour
are concerned with the modelling of chaotic systems
of continuous spectrum by means of discretised ones.
More precisely, the Kolmogorov-Sinai (KS) entropy of continuous
and discrete chaotic systems tend to coincide for a certain finite time range.
In this sense, the KS-entropy represents a robust indicator in the field
\cite{Tab79,Wal82,Awr16,Lat00,Tir01,Cas05,Fal14,Mih14,GomLya17}.
The coarse-graining of the quantum phase space,
as a consequence of the Uncertainty Principle (UP),
has an intimate relationship with quantum chaos time scales
\cite{Ike93,Eng97,Jaq05,Ino08,Gom17},
and quantum extensions for the KS-entropy
have been proposed \cite{Ben04,Cri93,Fal03}.

Since the quantum chaos time scales must be compatible with the imitation
of statistical properties of chaotic quantum systems in terms
of discretised models, then the type of statistics for such descriptions
should turn out to be relevant.
Tsallis nonextensive statistics is able to model KS-entropy
in a generalised way, from both regular and chaotic dynamics \cite{Tsa97}.
Nonextensive statistics has been applied to a wide variety
of systems and formalisms:
structures in plasmas \cite{Guo13},
entangled systems \cite{Kim16},
relativistic formulations \cite{Oli16},
quantum tunnelling and chemical kinetics \cite{Aqu17},
mathematical structures \cite{Nivanen03,Bor04,Lob09,Tem11}.
Many more examples can be found on \cite{tsallis-book}.
Numerical evidences of non-Boltzmannian chaotic behaviour
have been reported in low-dimensional conservative systems
\cite{Tir16,ruiz-et_al-jsm-2017,ruiz-et_al-pre-2017}
and dissipative ones \cite{Tir09,tirnakli-tsallis-2016}.
Some connections between quantum chaos and the nonextensivity formalism
has been advanced \cite{Tsa02},
but developments in relation to the characteristic time scales
in a general way seem to be still absent.

The goal of this paper is to provide a generalisation of
the quantum chaos time scales derived from nonextensive statistics.
They contain the relaxation and the random time scales as particular cases.
In addition, we also show how the generalised time scale
can arise as a consequence of an extended version of the KS-entropy
and the graininess of the quantum phase space.
The paper is organised as follows.
In Section \ref{sec:preliminaries}
we provide the preliminaries about the used formalism.
Section \ref{sec:quantum} is devoted to a brief review of
the relaxation and the random quantum chaos time scales.
In Section \ref{sec:generalized-time-scales},
we propose a generalised time scale
that has the relaxation and random time scales as special cases,
through a generalised KS-entropy and an asymptotical deformed uncorrelation.
We also address the fidelity decay,
and illustrate the generalised time scale with a kicked rotator with absorption.
Next, in Section \ref{sec:time-domain},
we give a discussion about a possible unified scenario
in the time domain of quantum chaos.
Finally, in Section \ref{sec:conclusions}
we draw our conclusions and future directions are outlined.

\section{\label{sec:preliminaries}Preliminaries}

In the following we give the necessary elements for the development
of the forthcoming sections.

\subsection{Kolmogorov-Sinai entropy}

\begin{figure}[htb]
 \begin{center}
  \includegraphics[width=0.45\textwidth,clip=]{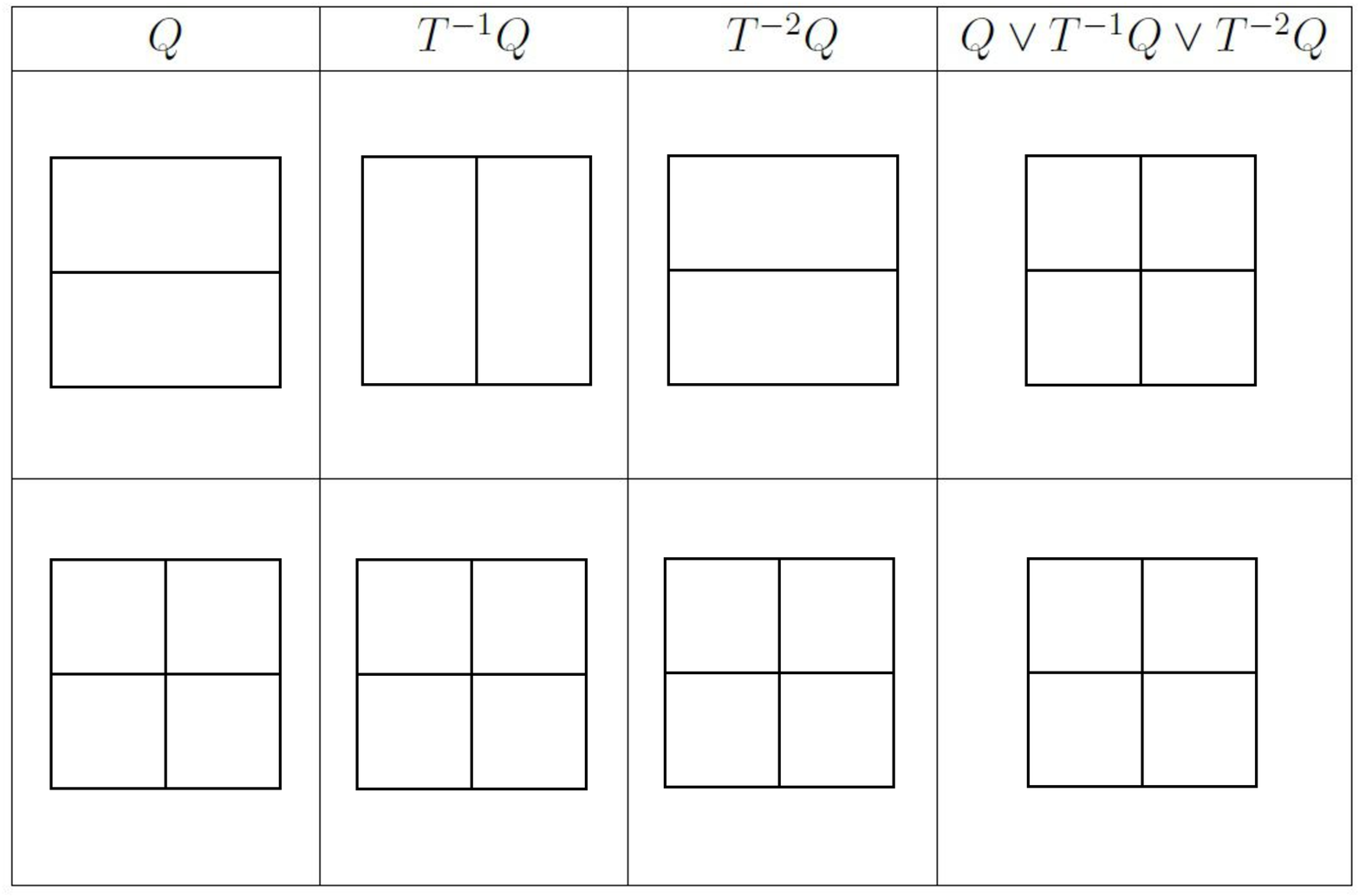}
 \end{center}
 \caption{\label{fig:refinement}
  An illustration of the refinement partition $\vee_{j=0}^n T^{-j} Q$
  for $n=2$, $\Gamma=[0,1]\times[0,1]$ and $T(x,y)=(y,-x)$, the $\pi/2$-rotation.
  Due to the rectangular symmetry of $T$
  (first line),
  at finite steps the initial rectangular partition is recovered,
  i.e., $Q=T^{-2}Q$.
  Thus, for this elementary and merely illustrative example, $h_{KS}(T)=0$.
 }
\end{figure}

A dynamical system is any quartern of the form
$(\Gamma, \Sigma, \widetilde{\mu}, \{T_t \}_{t\in J})$,
where $\Gamma$ is the phase space, $\Sigma$ is a $\sigma$-algebra,
$\widetilde{\mu}:\Gamma \rightarrow [0, 1]$ is a normalised measure and
$\{T_t \}_{t\in J}$ is a group%
\footnote{In some cases it is a semigroup,
          e.g., in discrete dynamical systems.}
of preserving measure transformations%
\footnote{For instance, in classical mechanics $T_t$ is the Liouville transformation.},
$J$ is typically the set of real numbers $\mathbb{R}$ for continuous dynamical systems
and the integers $\mathbb{Z}$ for discrete ones.
Dividing the phase space $\Gamma$ in $m$ small cells
$A_i$ of measure $\widetilde{\mu}(A_i)$,
the entropy of the partition $Q$ is
\begin{eqnarray}\label{KS entropy}
 H(Q)=\sum_{i=1}^{m}\widetilde{\mu}(A_i)\log \frac{1}{\widetilde{\mu}(A_i)}.
\end{eqnarray}
Given two partitions $Q_1$ and $Q_2$, it is possible to obtain
the refinement partition
$Q_1 \vee Q_2$:
$\{a_i \cap b_j : a_i \in Q_1 , b_j \in Q_2 \}$
($Q_1 \vee Q_2$ is a refinement of $Q_1$ and $Q_2$).
Starting from an arbitrary partition $Q$ of $\Gamma$,
the entropy of the refinement partition $H(\vee_{j=0}^n T^{-j} Q)$ can be derived,
with $T^{-j}$ the inverse of $T_j$ ($T^{-j} \equiv (T_j)^{-1}$),
and $T^{-j} Q = \{T^{-j} a : a \in Q\}$.
Fig. \ref{fig:refinement} depicts an elementary instance
of how the refinement partition is constructed
when the transformation is the $\pi/2$-rotation.
Each column refers to a time-step (backwards from left to right)
and each line refers to the graining
(increasing number of elements of the partition from up to down).

\noindent The standard measure theory defines the KS-entropy as \cite{Wal82}:
\begin{eqnarray}
 \label{entropy partition}
 h_{\textrm{\scriptsize KS}}(T) = \sup_{Q} \{\lim_{n\rightarrow\infty}\frac{1}{n}
 H(\vee_{j=0}^n T^{-j} Q)\},
\end{eqnarray}
where the supreme is taken over all measurable initial partitions $Q$
of $\Gamma$.
The definition of $h_{\textrm{\scriptsize KS}}(T)$
remains the same by replacing $T^{-j}$ by $T^{j}$;
we adopt $T^{-j}$ as is commonly used in the literature.
The KS-entropy
is the supreme of all the entropies per time step, corresponding to
each manner of dividing the phase space
(by refinement of the partitions evolved backwards in time)
allowed by the dynamics, when the time step tends to infinity.
Due to the rectangular symmetry of the time evolution operator $T$
of the example of Fig. \ref{fig:refinement},
any rectangular partition
$Q=\{A_1,\ldots,A_m\}$ with all the $A_i$ of the same volume
$\widetilde{\mu}(A_i)=1/m$ and
$m=4^{l}$ ($l\in\mathbb{N}$, $l\geq1$) will be invariant after $T^{-1}$,
i.e. $Q=T^{-1}Q$
(second line of Fig. \ref{fig:refinement}, $l=1$).
So, $\vee_{j=0}^n T^{-j} Q=Q$ for all $n\in\mathbb{N}$
and then using the fact that $T$ preserves the measure we obtain
$H(\vee_{j=0}^n T^{-j} Q)=H(Q)=\log m$ for all $n\in\mathbb{N}$.
Hence, since the limit
$\lim_{n\rightarrow\infty}\frac{1}{n}H(\vee_{j=0}^n T^{-j} Q) =
 \lim_{n\rightarrow\infty}\frac{\log m}{n} = 0$,
then taking the supreme over rectangular partitions $Q$
(having $m=4^l$ elements with $l\rightarrow\infty$),
it follows that $h_{\textrm{\scriptsize KS}}(T)=0$.
The same argument can be extended for all $\omega \pi$-rotation
with rational $\omega$.

In addition, in the context of information
theory the Brudno theorem expresses that
the KS-entropy is the average unpredictability of information
of all possible
trajectories in the phase space.
In turn, Pesin theorem relates the KS-entropy with the exponential
instability of motion given by the Lyapunov exponents,
thus leading $h_{\textrm{\scriptsize KS}}>0$ as a sufficient condition
for the chaotic motion%
\footnote{%
 Distinguishable levels within the chaotic behaviour
 can occur like hyper-chaotic, hyper-hyper chaotic,
 etc.\ \protect\cite{Awr16}.
         }.

If the dynamical system has a characteristic time scale
expressed by a dimensionless parameter $\kappa$,
it follows from (\ref{KS entropy}) and (\ref{entropy partition})
that $h_{\textrm{\scriptsize KS}}$ can be written as
(see Theorem 4.13 of \cite{Wal82})
\begin{eqnarray}
 \label{rescaled KS entropy}
 h_{\textrm{\scriptsize KS}}(T)
 = \frac{1}{\kappa}h_{\textrm{\scriptsize KS}}(T_{\kappa}),
\end{eqnarray}
where $h_{\textrm{\scriptsize KS}}(T_\kappa)$ is obtained from (\ref{KS entropy})
replacing $T$ by $T_{\kappa}$,
and $T_{\kappa}$ is the transformation $T_t$ for $t=\kappa$.
The time rescaling allows to connect
the characteristic time scale $\kappa$ and the KS-entropy by means
of (\ref{rescaled KS entropy}).
This concept will be relevant for obtaining the generalised time scale.

\subsection{Deformed quantities from nonextensive statistical mechanics}

Motivated by the generalised $q$-entropy \cite{Tsa88}
\begin{eqnarray}\label{tsallis entropy}
 S_q=k \frac{\sum_{i=1}^{W}p_i^q-1}{1-q}
\end{eqnarray}
($q \in \mathbb{R}$ is the entropic index,
 $k$ is a positive constant that defines the unity in which $S_q$ is measured,
 and $(p_1,\ldots,p_W)$ is a discretised probability distribution),
the $q$-logarithm and the $q$-exponential functions
(one is inverse of the other) are defined \cite{Tsa94}:
\begin{eqnarray}
 \label{q-functions}
 \begin{array}{lll}
  \ln_{q}x &=& \frac{x^{1-q}-1}{1-q} \quad (x>0),\\
  e_{q}(x) &=& [1+(1-q)x]_{+}^{\frac{1}{1-q}} \quad (x\in\mathbb{R}),
 \end{array}
\end{eqnarray}
where $[A]_+ =\max\{A,0\}$.
The $q$-entropy is then rewritten as
\begin{eqnarray}
 \label{sq-with-qlog}
 S_q=k \sum_{i=1}^{W} p_i \ln_q (1/p_i).
\end{eqnarray}
It is straightforward to verify that
\begin{eqnarray}
 \label{q-functions properties}
 \begin{array}{lll}
  \ln_{q}(xy)      &=&\ln_{q}x+\ln_{q}y+(1-q)\ln_{q}x\ln_{q}y.
 \end{array}
\end{eqnarray}
This expression justifies $S_q$ to be referred to as
\textit{nonadditive entropy}:
the $q$-entropy of a system composed by two independent subsystems $A$ and $B$
($p_{ij}(A+B) = p_i(A) p_j(B)$)
is
$ S_q(A+B) = S_q(A) + S_q(B) + \frac{1-q}{k} S_q(A) S_q(B) $.
Equation (\ref{q-functions properties}) is one of the relations that triggers
the generalisation of the usual algebraic operations \cite{Nivanen03,Bor04}:
\begin{eqnarray}\label{q-algebra}
 \begin{array}{lll}
  x\oplus_{q}  y &=& x+y+(1-q)xy, \\
  x\ominus_{q} y &=& \frac{x-y}{1+(1-q)y} \quad (y\neq \frac{1}{q-1}),\\
  x\otimes_{q} y &=& [x^{1-q}+y^{1-q}-1]_{+}^{\frac{1}{1-q}} \quad(x,y>0), \\
  x\oslash_{q} y &=& [x^{1-q}-y^{1-q}+1]_{+}^{\frac{1}{1-q}} \quad (x,y>0),
 \end{array}
\end{eqnarray}
called $q$-sum, $q$-difference, $q$-multiplication, and $q$-division,
respectively.
One expression that stems from these generalised algebraic relations,
and that plays a central role in the development to come, is
\begin{eqnarray}
 \label{eq:ln_q-qproduct}
 \ln_q( x \otimes_q y ) = \ln_q x + \ln_q y.
\end{eqnarray}

\section{\label{sec:quantum}Quantum chaos time scales}

Some classical chaos conditions, like a continuous spectrum and phase space,
are difficult to be used for defining quantum chaos \cite{Cas95}.
These conditions are frequently violated in quantum mechanics,
because the majority of quantum systems of interest have discrete spectrum,
and according to the UP, the phase space must be discretised by cells
of finite size $\Delta x\Delta p\geq h$ (per degree of freedom)
with $h$ the Planck constant.
The situation becomes even more tricky in relation to the CP,
since it prescribes that all classical phenomena (including chaos)
are expected to emerge from the underlying quantum domain in the classical limit,
when the Planck constant is vanishingly small.
A study of this issue can be found on \cite{Ike93,Eng97}.
These facts stimulate the search for a quantum formalism
that is consistent with the UP and the CP,
and that also allow to explain the emergence of chaos in the classical limit.
A granulated (discretised) phase space for quantum dynamics given by the UP
appears to be a suitable treatment of the problem.
However, even taking into account the graininess,
some complications arise when attempting to make a quantisation
of a chaotic system.
For instance, the compactness of a chaotic phase space
leads to a discrete spectrum.
Thus, the task is subtle and requires an adequate tool
that can capture the main dynamical properties
with respect to the continuous spectrum of the chaotic systems.
As mentioned in the introduction, the KS-entropy constitutes a good candidate
due to its robust signature, both in theory and applications,
in the modelling of classical chaotic systems from discretised ones.
The KS-entropy is equal to the Shannon entropy per time step of the ensemble
of the trajectory bunches in the limit of infinitely many time steps,
and therefore represents an information measure
of the dynamical system in the asymptotic limit.
This feature is expressed in a compact form by the Pesin theorem,
that links the KS-entropy with the Lyapunov coefficients.
In this regard, for a description of quantum systems exhibiting chaotic behaviour,
the desirable would be a quantum extension of the KS-entropy.
Several non-commutative candidates has been proposed,
where the KS-extensions yield the classical KS-entropy within finite
time ranges.
This fact is asserted by some authors as the singular one par excellence
in quantum chaos \cite{Ben04,Cri93,Fal03}.

Two time scales characterising the quantum motion, regular and chaotic,
are well distinguished:  the relaxation  time scale $\tau_{R}$,
and the random time scale $\tau_{r}$.
For the regular case, the classical and quantum behaviours
approximate each other for
\begin{eqnarray}
 \label{power law time scale}
 t \leq\tau_{R}, \quad \textrm{with} \quad
 \tau_{R}\propto \eta^{\alpha},
\end{eqnarray}
where $\eta=\frac{I}{h^D}$ is the quasiclassical parameter
of the order of the characteristic value of the classical action $I$,
$2D$ is the dimension of the phase space,
and $\alpha>0$ is a system-dependent parameter.
The relaxation time establishes the so-called semiclassical regime,
with the particularity that within this one the discrete spectrum
cannot be solved if $t \leq \tau_{R}$
(see p.\ 12 of \cite{Cas95}).
On the other hand, the random time scale $\tau_{r} \ll \tau_{R}$ determines
the time interval where the wave packet motion spreads over the phase space
and it is related to a property of the strong chaos,
the exponential instability,
that is measured by the positive Lyapunov exponents.
This is given by
\begin{eqnarray}
 \label{logarithmic time scale}
 t \leq\tau_r, \quad \textrm{with}
 \quad
 \tau_r=\frac{\ln \eta}{h_{\textrm{\scriptsize KS}}(T)},
\end{eqnarray}
where $h_{\textrm{\scriptsize KS}}(T)$ is the KS-entropy of the
classical analogue having the classical Liouville evolution $T$.
$\tau_r$ represents a resolution for the disagreement between
the CP and the classical limit,
and an evidence of the noncommutative double limit
\cite{Ben04,Cri93,Fal03}
\begin{eqnarray}
 \label{double limits}
 \lim\limits_{t\rightarrow\infty}\lim\limits_{\eta\rightarrow\infty}
  \neq
 \lim\limits_{\eta\rightarrow\infty}\lim\limits_{t\rightarrow\infty},
 \nonumber
\end{eqnarray}
where the left hand side means classical chaos
while the right hand side expresses a quantum behaviour without chaos
(see p.\ 17 of \cite{Cas95}).

\section{\label{sec:generalized-time-scales}%
        Generalised time scale}

The $q$-logarithm asymptotically behaves as a power law for $q\neq 1$
(equation\ (\ref{q-functions})),
and recovers the natural logarithm when $q \rightarrow 1$.
The crucial observation is that these two behaviours exactly coincide
with the relaxation and random times with respect to the quasiclassical
parameter (equations\ (\ref{power law time scale})
and (\ref{logarithmic time scale})).
In order to unify these two time scales,
the first step is to generalise the KS-entropy.
Generalised versions of the KS-entropy have been proposed
from different contexts \cite{Lat00,Tir01,Cas05, Fal14,Mih14}.
Stimulated by the relation between the KS-entropy and
the random time scale, equation\ (\ref{logarithmic time scale}),
we make the reasonable assumption that the same could be expected to happen
between a generalisation of the KS-entropy and the corresponding time scale.
Our strategy is in connection with the generalised
KS-entropies studied by \cite{Fal14} and supported numerically by \cite{Cas05}.
The entropy partition $H(Q)$ can be deformed as
(see equation\ (\ref{sq-with-qlog}), with $k=1$)
\begin{eqnarray}
 \label{deformed entropy partition}
 H_{q}(Q)=\sum_{i=1}^{m}\widetilde{\mu}(A_i)\ln_{q}\frac{1}{\widetilde{\mu}(A_i)}
\end{eqnarray}
($H_q(Q) \ge 0, \forall q$),
which reduces to the usual one when $q\rightarrow1$.
We propose the
\emph{generalised} KS-entropy as
\begin{eqnarray}
 \label{deformed KS entropy}
 h_{\textrm{\scriptsize KS}}^{(q)}(T) =
   \sup_{Q} \{
              \lim_{n\rightarrow\infty}
              \frac{1}{n} H_{q}(\vee_{j=0}^n T^{-j} Q)
            \}.
\end{eqnarray}
The scaling property given by equation\ (\ref{rescaled KS entropy}) is still valid:
\begin{eqnarray}
 \label{deformed rescaled KS entropy}
 h_{\textrm{\scriptsize KS}}^{(q)}(T)
 =
 \frac{1}{\kappa}h_{\textrm{\scriptsize KS}}^{(q)}(T_{\kappa}).
\end{eqnarray}
The random time scale can be readily generalised into
\begin{eqnarray}
 \label{q--time scale}
  \tau_q =
  \displaystyle
  \frac{\ln_q \eta}{h_{\textrm{\scriptsize KS}}^{(q)}(T)}
  \propto \ln_q \eta, \quad \eta\geq1,
\end{eqnarray}
where $q$ is the entropic index,
$T$ is the classical Liouville evolution,
and $h_{\textrm{\scriptsize KS}}^{(q)}(T)$
is the generalisation of the KS-entropy.
The condition $\eta \geq 1$ ensures that $\tau_q$ is nonnegative, as expected.
Physically, $\eta \geq 1$ says that the classical action $I \ge h^D$,
which includes the region of the classical limit.
The relaxation and random times scales are particular cases of $\tau_q$:

\noindent \emph{Relaxation time scale}.
If $q=1-\alpha$ in equation\ (\ref{q--time scale}),
then using (\ref{q-functions}) in the classical limit $\eta \gg 1$,
\begin{eqnarray}
 \label{q--time scale Heisenberg}
  \tau_{q=1-\alpha} =
  \frac{\ln_{1-\alpha} \eta}{h_{\textrm{\scriptsize KS}}^{(1-\alpha)}(T)}
  =
  \frac{\eta^\alpha-1}{\alpha h_{\textrm{\scriptsize KS}}^{(1-\alpha)}(T)}
  \approx
  \frac{\eta^\alpha}{\alpha h_{\textrm{\scriptsize KS}}^{(1-\alpha)}(T)}
  \propto \eta^\alpha,
\end{eqnarray}
that is precisely the relaxation time scale $\tau_R$.

\noindent \emph{Random time scale}.
This time scale is directly obtained within the proper limit:
\begin{eqnarray}
 \label{q--time scale Ehrenfest}
  \lim\limits_{q \rightarrow 1} \tau_q =
  \lim\limits_{q \rightarrow 1}
  \frac{\ln_q \eta}{h_{\textrm{\scriptsize KS}}^{(q)}(T)}
  =
  \frac{\ln \eta}{h_{\textrm{\scriptsize KS}}(T)}.
\end{eqnarray}

\subsection{\label{subsec:graininess}Graininess in a deformed decay scenario}

We provide a justification for how could arise the generalised time scale
$\tau_q$ from the point of view of $h_{\textrm{\scriptsize KS}}^{(q)}(T)$.
In fact, $h_{\textrm{\scriptsize KS}}^{(q)}(T)$ allows one to obtain
$\tau_q$ by considering the graininess
of the phase space, due to the UP, as follows.
We consider a quantum system of $D$ degrees of freedom
so the discretised quantum phase is $2D$-dimensional
and coarse-grained by
undeformable cells of minimal size $\Delta q \Delta p = h^{D}$,
where $(q,p)$ denotes $(q_1,\ldots,q_D,p_1,\ldots,p_D)$.
The motion of the system lies on a
bounded compact region $\Omega\in\mathbb{R}^{2D}$,
where its Lebesgue measure
$\mu(\Omega)<\infty$ is of the order of the classical action $I$,
i.e. (see equation\ (\ref{power law time scale})),
\begin{eqnarray}
 \label{eq:eta}
 \eta = \frac{\mu(\Omega)}{h^{D}}.
\end{eqnarray}
This presumption appropriately meets in chaotic billiards%
\footnote{
 Since closed systems with $D=1$ are integrable, and therefore not chaotic.
         }
with $D>1$,
or in non integrable systems under a central potential
(for example, the Henon-Heiles system).
The UP implies that there exists a maximal partition
(that is, the greatest refinement that one can take)
$Q_{\scriptsize{\textrm{max}}}=\{A_1,\ldots,A_M\}$ of $\Omega$ constituted by
$M$ identical and rigid rectangle cells $A_i$ of dimensions $\Delta q\Delta p$
and a dimensionless normalised measure
$\widetilde{\mu}(A_i)=\frac{h^{D}}{\mu(\Omega)}$ for all $i=1,\ldots,M$
$(\widetilde{\mu}= \frac{\mu}{\mu(\Omega)})$,
where $M$ is the maximal number of cells $A_i$ contained in $\Omega$.
$Q_{\scriptsize{\textrm{max}}}$ is a partition, so
$\sum_{i=1}^M \widetilde{\mu}(A_i)=\sum_{i=1}^M\frac{h^{D}}{\mu(\Omega)}=1$,
what implies
\begin{equation}
 \label{graininess}
 M h^{D}=\mu(\Omega),
\end{equation}
that is simply an expression of the \emph{graininess}
of the quantum phase space ($M = \eta$).
Next step is to perform the $h_{\textrm{\scriptsize KS}}^{(q)}(T_\kappa)$
assuming that the system has a finite time scale $\kappa$ in which the classical
and quantum descriptions tend to coincide.
Therefore, the supreme in (\ref{deformed KS entropy}) can be replaced by
$\lim_{n\rightarrow\infty}
 \frac{1}{n}H_q(\vee_{j=0}^{n}T_{\kappa}^{-j}Q_{\scriptsize{\textrm{max}}})$
in the context of the graininess.
It is a difficult task to calculate
\begin{eqnarray}
 \label{supreme deformed entropy}
  H_q(\vee_{j=0}^{n}T_{\kappa}^{-j}Q_{\scriptsize{\textrm{max}}})
  =
  \sum_{(i_0,i_1,\ldots,i_n)}
  \widetilde{\mu}(\cap_{j=0}^{n} T_{\kappa}^{-j}A_{i_j})\:
  \ln_{q}\frac{1}{\widetilde{\mu}(\cap_{j=0}^{n} T_{\kappa}^{-j}A_{i_j})},
\end{eqnarray}
due to the form that the elements of
$\vee_{j=0}^{n}T_{\kappa}^{-j}Q_{\scriptsize{\textrm{max}}}$
can adopt as $n$ increases up to infinity.
However, an assumption can be made to overcome this.
Typically, the dynamics in chaotic systems is such that
a decay of correlations between subsets of phase space
sufficiently separated in time
is expected to take place in the asymptotic limit
(for instance, in mixing systems).
The information about the correlation decay is precisely contained
in the way in which
$\widetilde{\mu}(\cap_{j=0}^{n} T_{\kappa}^{-j}A_{i_j})$
decreases as $n \to \infty$ \cite{Lich92}.
Each set $\cap_{j=0}^{n} T_{\kappa}^{-j}A_{i_j}$
corresponds to a part of the phase space $\Gamma$
which is divided through the refinement of all the partitions
that result from the maximal one, $Q_{\scriptsize{\textrm{max}}}$,
evolved backwards up to $j$-th time-step
(i.e. $T_{\kappa}^{-j}Q_{\scriptsize{\textrm{max}}}$).
Thus, for all $(n+1)$-tuple $(i_0,i_1,\ldots,i_n)$ of $n+1$ labels in
$\{1,\ldots,M\}$, the set
$\cap_{j=0}^{n} T_{\kappa}^{-j}A_{i_j}$
is the central object
for studying the dynamics, from the point of view of the KS-entropy.

We conjecture that, in the asymptotical regime,
\begin{eqnarray}
 \label{uncorrelation deformed entropy}
  \displaystyle
  \frac{1}{\widetilde{\mu}(\cap_{j=0}^{n} T_{\kappa}^{-j}A_{i_j})}
  =
  \displaystyle
  \frac{1}{\widetilde{\mu}( A_{i_0})} \otimes_q
  \frac{1}{\widetilde{\mu}( T_{\kappa}^{-1}A_{i_1})}
  \otimes_q \ldots \otimes_q \frac{1}{\widetilde{\mu}( T_{\kappa}^{-n}A_{i_n})}
\end{eqnarray}
for all positive integer $n$, where
$\frac{1}{\widetilde{\mu}(A_{i_0})}
 \otimes_q
 \frac{1}{\widetilde{\mu}(T_{\kappa}^{-1}A_{i_1})}
 \otimes_q \ldots \otimes_q
 \frac{1}{\widetilde{\mu}( T_{\kappa}^{-n}A_{i_n})}$
stands for the $q$-product between
$\frac{1}{\widetilde{\mu}( T_{\kappa}^{-j}A_{i_j})}$ from $j=0$ up to $n$.

The $q$-product of $\frac{1}{\widetilde{\mu}( T_{\kappa}^{-n}A_{i_j})}$
introduces a special and particular correlation that leads to the
generalised time scale according to equation\ (\ref{q--time scale}).
The particular case $q=1$ corresponds to the simple product of probabilities
of Bernoulli dynamical systems, typical for uncorrelated systems.
In view thereof,
equation\ (\ref{uncorrelation deformed entropy}) represents
an \emph{asymptotical deformed uncorrelation}
between the subsets $T_{\kappa}^{-j}A_{i_j}$,
a wider condition than the \emph{mixing of all orders}%
\footnote{Note that only mixing does not guarantees the existence of positive
          Lyapunov exponents and therefore neither the condition
          $h_{\textrm{\scriptsize KS}}>0$.
          Instead, the condition of mixing of all orders
          $\lim_{n_0,\ldots,n_k\rightarrow\infty}
                 \mu(A_0 \cap T^{-n_1}A_{1} \cap \ldots \cap T^{-n_k}A_{k})
               = \mu(A_0)\mu(A_1)\cdots\mu(A_k)$
          allows to treat the sum involved in $h_{\textrm{\scriptsize KS}}$.},
that happens for $q \to 1$.
$\widetilde{\mu}$ is preserved%
\footnote{
 Ergodic theory focuses on dynamical systems with invariant densities
 that are preserved for some function.
 In classical mechanics, the Liouville theorem expresses this condition,
 through the conservation of the volume $\mu$ of the phase space.
         },
inasmuch as $T_t$ preserves $\mu$, hence
\begin{eqnarray}
 \label{measure preserving 1}
 \widetilde{\mu}( T_{\kappa}^{-j}A_{i_j})
 =
 \widetilde{\mu}( A_{i_j})
 =
 \frac{1}{\eta},
 \quad \forall \ j=0,\ldots,n.
\end{eqnarray}
From equations\ (\ref{eq:ln_q-qproduct}) and (\ref{measure preserving 1}),
the deformed uncorrelation condition can be written as
\begin{eqnarray}
 \label{uncorrelation deformed entropy 2}
  \frac{1}{\widetilde{\mu}(\cap_{j=0}^{n} T_{\kappa}^{-j}A_{i_j})}
  =
  e_q^{ (n+1) \ln_q \eta }.
\end{eqnarray}
To complete the calculus it is sufficient to replace
(\ref{uncorrelation deformed entropy 2}) in (\ref{supreme deformed entropy})
to obtain
\begin{eqnarray}
 \label{deformed entropy calculated}
 \begin{array}{lll}
  \displaystyle
  H_q(\vee_{j=0}^{n}T_{\kappa}^{-j}Q_{\scriptsize{\textrm{max}}})
  &=&
  \displaystyle
  \sum_{(i_0,i_1,\ldots,i_n)}
  \widetilde{\mu}(\cap_{j=0}^{n} T_{\kappa}^{-j}A_{i_j})
  (n+1) \ln_q \eta
  \\
  &=&
  (n+1) \ln_q \eta,
 \end{array}
\end{eqnarray}
where
$\sum_{(i_0,i_1,\ldots,i_n)}
 \widetilde{\mu}(\cap_{j=0}^{n} T_{\kappa}^{-j}A_{i_j})=1$
due to the fact that
$\vee_{j=0}^{n}T_{\kappa}^{-j}Q_{\scriptsize{\textrm{max}}}$
and $Q_{\scriptsize{\textrm{max}}}$
have the same measure, since the former is a refinement of the latter,
and the motion is confined to $\Omega$.
Then, from (\ref{deformed entropy calculated}),
\begin{eqnarray}
 \label{deformed KS entropy calculated 2}
 \begin{array}{lll}
  h_{\textrm{\scriptsize KS}}^{(q)}(T_\kappa)
  &=&
      \displaystyle
      \lim_{n\rightarrow\infty}
      \textstyle
      \frac{1}{n}H_q(\vee_{j=0}^{n}T_{\kappa}^{-j}Q_{\scriptsize{\textrm{max}}})
  \\
  &=&
      \displaystyle
      \lim_{n\rightarrow\infty}
      \textstyle
      \left( \frac{n+1}{n} \right) \ln_q \eta
  \\
  &=& \ln_q \eta,
 \end{array}
\end{eqnarray}
and from (\ref{deformed rescaled KS entropy}),
\begin{eqnarray}
 \label{deformed KS entropy compact expression}
 h_{\textrm{\scriptsize KS}}^{(q)}(T_\kappa)
 = \ln_q \eta
 = \kappa \, h_{\textrm{\scriptsize KS}}^{(q)}(T).
\end{eqnarray}
Finally, it follows that
\begin{eqnarray}
 \label{deformed time}
 \kappa = \frac{1}{h_{\textrm{\scriptsize KS}}^{(q)}(T)}
          \ln_q \eta
\end{eqnarray}
which is the generalised time scale $\tau_q$.

With the help of the $q$-product definition (see equation\ (\ref{q-algebra})),
the asymptotical deformed uncorrelation,
equation\ (\ref{uncorrelation deformed entropy}),
can be expressed as
\begin{eqnarray}
 \label{role uncorrelation deformed}
 \frac{1}{\widetilde{\mu}(\cap_{j=0}^{n} T_{\kappa}^{-j}A_{i_j})}
 =
 \left(
  \sum_{j=0}^{n}\left(
                      \frac{1}{\widetilde{\mu}(T_{\kappa}^{-j}A_{i_j})}
                \right)^{1-q}- n
 \right)^{\frac{1}{1-q}}.
\end{eqnarray}
Considering equations\ (\ref{graininess}) and (\ref{measure preserving 1}),
it results
\begin{eqnarray}
 \label{main deformed uncorrelation}
 \frac{1}{\left[
                \widetilde{\mu}(\cap_{j=0}^{n} T_{\kappa}^{-j}A_{i_j})
          \right]^{1-q}}
 =
 (n+1)M^{1-q}-n,
\end{eqnarray}
where $M>1$ is a positive integer.
This equation is the starting point for characterising
some typical correlation decays.

\subsubsection{Extensive case $q\rightarrow 1$: chaotic dynamics}.
It is easy to see that when the entropic index $q$ tends to one,
the formula (\ref{main deformed uncorrelation}) is trivially satisfied.
Then, for all value of $M$ the asymptotical uncorrelation holds,
where $M$ is the order of the quasiclassical parameter $\eta$.
This means that for a chaotic dynamics, governed by the random time scale,
the asymptotical uncorrelation is valid for all values of $M$.

\subsubsection{Nonextensive case $q=1-\alpha < 1$:
regular and non-chaotic dynamics}.
One interesting case occurs when $q=1-\alpha$
in (\ref{main deformed uncorrelation}), thus obtaining
\begin{eqnarray}
 \label{main deformed uncorrelation regular}
 \widetilde{\mu}(\cap_{j=0}^{n} T_{\kappa}^{-j}A_{i_j})
 = \left(\frac{1}{(n+1)M^{\alpha}-n}\right)^{\frac{1}{\alpha}}.
\end{eqnarray}
In turn, this equation can be approximated, using the binomial formula
$(1-\varepsilon)^{\gamma}\simeq 1-\gamma\varepsilon$, as
\begin{eqnarray}
 \label{main deformed uncorrelation regular approx}
 \widetilde{\mu}(\cap_{j=0}^{n} T_{\kappa}^{-j}A_{i_j})
 \simeq
 \frac{1}{M(n+1)^{\frac{1}{\alpha}}}
 \propto (n+1)^{-\frac{1}{\alpha}}
\end{eqnarray}
in the classical limit ($M \gg 1$).
Equation\ (\ref{main deformed uncorrelation regular approx}) expresses that
the correlation decay follows a power law,
which is precisely the observed in regimes that are not completely chaotic
but provided with a complex dynamics \cite{Cos97}.
The volume of the sets
$\cap_{j=0}^{n} T_{\kappa}^{-j}A_{i_j}$
can take values less than $h^D$
in the asymptotic limit of sufficiently large $n$,
a particular feature that can occur in mixed regimes
of classically chaotic systems \cite{Zur01}.
These type of regions in the phase space are called
\emph{sub-Planck structures},
and they were studied in decoherence phenomena.

\subsection{\label{subsec:fidelity}
Fidelity decay}

The fidelity was introduced by Peres \cite{Per95}
as an indicator for the stability of quantum motion.
It plays a role analogous to the Lyapunov exponents in classical
mechanics. In this Section we generalise this concept
in view of making it coherent with the generalised time scale.
Through a perturbation $\widehat{V}$ on a Hamiltonian $\widehat{H}_0$,
$\widehat{H}=\widehat{H}_0+\widehat{V}$
with an arbitrary initial state $|\psi_0\rangle$,
Peres defined fidelity as
the overlap of a state at time $t$ with its perturbed echo:
\begin{eqnarray}
 \label{fidelity}
 M(t)=|\langle \psi_0 |\widehat{U}^{\prime}_{-t} \widehat{U}_{t} |\psi_0\rangle |^2,
\end{eqnarray}
the so-called \emph{Loschmidt echo}.
The unperturbed and the perturbed evolution operators are
$\widehat{U}_{t}=\exp(-i\widehat{H}_0t)$ and $\widehat{U}^{\prime}_{-t}=\exp(i\widehat{H}t)$,
respectively.
The perturbed evolution operator $\widehat{U}^{\prime}_{t}$
can be expressed in a more convenient form:
\begin{eqnarray}
 \label{perturbedevoution}
 \widehat{U}^{\prime}_{t} = \widehat{U}_{\delta}^{t}
                      =\widehat{U}_{t} \exp(-i\widehat{B}\delta/\hbar),
\end{eqnarray}
where $\widehat{B}$ is a self-adjoint operator,
$\delta$ is the perturbation strength
and $\hbar$ is an effective Planck constant \cite{Jal01}.
Analytical evaluation of equation\ (\ref{fidelity}) may be difficult
due to the non-commutativity between the unperturbed Hamiltonian $\widehat{H}_0$
and the perturbation $\widehat{B}$.
This difficulty is analogous to that of evaluating the KS-entropy
(see equation\ (\ref{supreme deformed entropy})):
in one case, the hindrance lies on
the product of non-commutative operators,
and in the other,
on the intersection of elements of different partitions of the phase space.
To overcome this, the fidelity may be expanded in power series
of some characteristic parameter.
The Loschmidt echo is obtained from
equations\ (\ref{fidelity}) and (\ref{perturbedevoution})
as a power series of the perturbation strength $\delta$ \cite{Pro02}:
\begin{eqnarray}
 \label{fidelity-series}
 M(t)=\left|
            1+\sum_{m=1}^{\infty}\frac{i^m \delta^m}{m!\hbar^m}
            \widehat{\mathcal{T}}\sum_{t_1,\ldots,t_m=0}^{t-1}
            \langle \widehat{B}_{t_1}\cdots\widehat{B}_{t_m} \rangle
      \right|^2,
\end{eqnarray}
where $\widehat{\mathcal{T}}$ stands for a left-to-right time ordering,
$\widehat{B}_{t_j}=\widehat{U}^{\prime}_{t_j}\widehat{B}\widehat{U}^{\prime}_{-t_j}$
represents the perturbation at time $t_j$ for all $j=1,\ldots,m$,
and $t$ is a discrete time in units of the period%
\footnote{Here we are assuming a periodically time-dependent Hamiltonian that
          represents a sufficiently large class of chaotic quantum systems,
          the Floquet representation.}
of the Hamiltonian $\widehat{H}$.
The Weyl-Wigner transform permits to recover the dynamics
of the quantum phase space in the classical limit \cite{Hil84,Gad95}.
Along this line,
the mean value
$\langle \widehat{B}_{t_1}\cdots\widehat{B}_{t_m}\rangle$,
that carries the time dependence of the fidelity decay,
may be expressed by
\begin{eqnarray}
 \label{meanvalue-B}
 \begin{array}{lll}
  \langle \widehat{B}_{t_1}\cdots\widehat{B}_{t_m} \rangle
  &=&
      \textrm{Tr} \left(
                   \displaystyle
                   \frac{1}{N}\widehat{1}\widehat{B}_{t_1}\cdots\widehat{B}_{t_m}
                  \right)
  \\
  &=&
  \displaystyle
      \frac{1}{N h^D}
      \int_{\mathbb{R}^{2D}}
           \textrm{\hugetilde{\widehat{B}_{t_1}\cdots\widehat{B}_{t_m}}}
            dq dp
 \end{array}
\end{eqnarray}
The classical chaotic dynamics is achieved by approximating
the Weyl symbol of the product of $\widehat{B}_{t_j}$
by the product of the Weyl symbols of each one:
\begin{eqnarray}
 \label{B-factorization}
 \textrm{\hugetilde{
            \widehat{B}_{t_1}\cdots\widehat{B}_{t_m}
           }}
  \simeq
  \widetilde{B}_{t_1}\cdots\widetilde{B}_{t_m}.
\end{eqnarray}
Then, equations\ (\ref{meanvalue-B}) and (\ref{B-factorization}) are rewritten as
\begin{eqnarray}
 \label{meanvalue-chaotic}
 \begin{array}{lll}
  \langle \widehat{B}_{t_1}\cdots\widehat{B}_{t_m} \rangle
  &\simeq&
  \displaystyle
      \frac{\mu(\Omega)}{N h^D}
      \int_{\Omega}
           \widetilde{B}_{t_1}\cdots\widetilde{B}_{t_m} d\widetilde{\mu},
 \end{array}
\end{eqnarray}
with
$d\mu = dq dp$,
and
$\widetilde{B}_{t_j}$ are
assumed to have a support in $\Omega$ for $t_m\rightarrow\infty$
i.e., the perturbation acts on $\Omega$ only.
The quantum interference correlations are cancelled
by the chaotic dynamics in phase space,
expressed mathematically by (\ref{meanvalue-chaotic}),
giving place to a coarse-grained distribution \cite{Cas95}, where
$Q_{\scriptsize{\textrm{max}}}=\{A_1,\ldots,A_M\}$
is the finest partition of $\Omega$.
As $M \to \infty$ in the classical limit,
$\widetilde{B}(q,p)$
becomes a function defined over
the granulated quantum phase space
that can be approximated as a linear
combination of the characteristic functions,
$1_{A_j}(q,p)$ for all $j=1,\ldots,M$:
\begin{eqnarray}
 \label{B-characteristic-function}
 \widetilde{B}(q,p)\simeq\sum_{j=1}^{M}\gamma_j 1_{A_j}(q,p),
\end{eqnarray}
then,
\begin{eqnarray}
 \label{meanvalue-product2}
   \langle \widehat{B}_{t_1} & \cdots \widehat{B}_{t_m} \rangle
  \simeq \nonumber \\
    &\frac{\eta}{N}
     \int_{\Omega}
       \left(
             \sum_{j_1=1}^{M} \gamma_{j_1} 1_{T_{t_1} A_{j_1}}(q,p)
             \cdots
             \sum_{j_m=1}^{M} \gamma_{j_m} 1_{T_{t_m} A_{j_m}}(q,p)
       \right)
       d\widetilde{\mu}
   \nonumber \\
    &\simeq \frac{\eta}{N}
      \sum_{j_1,\ldots,j_m=1}^{M}
           \gamma_{j_1} \cdots \gamma_{j_m}
           \widetilde{\mu}(T_{t_1} A_{j_1}\cap\ldots \cap T_{t_m} A_{j_m}).
\end{eqnarray}
At this point we introduce a hypothesis with the same structure
of equation\ (\ref{uncorrelation deformed entropy}) in order to make the
fidelity decay scenario compatible with the generalised time scale.
We replace $(\gamma_{j_1}\cdots\gamma_{j_m})^{-1}$ by
$(\gamma_{j_1})^{-1}\otimes_q\cdots\otimes_q(\gamma_{j_m})^{-1}$
in equation\ (\ref{meanvalue-product2}) so
$\langle \widehat{B}_{t_1}\cdots\widehat{B}_{t_m}\rangle$
is generalised as
\begin{eqnarray}
 \label{generalized-meanvalue-product}
 & \langle \widehat{B}_{t_1} \cdots \widehat{B}_{t_m}\rangle_q
  =\nonumber\\
 &\frac{\eta}{N}
  \sum_{j_1,\ldots,j_m=1}^{M}
 [(\gamma_{j_1})^{-1}\otimes_q\cdots\otimes_q(\gamma_{j_m})^{-1}]^{-1}
 \widetilde{\mu}(T_{t_1} A_{j_1}\cap\ldots \cap T_{t_m} A_{j_m}).
\end{eqnarray}
The following step is to take a uniformly distributed perturbation
over $\Omega$,
$\gamma_j=\frac{1}{M}$ for all $j$
 (the unit of the perturbation strength $\delta$
  may be properly chosen to turn $\widehat{B}$ dimensionless),
so
$(\gamma_{j_1})^{-1}\otimes_q\cdots\otimes_q(\gamma_{j_m})^{-1}
= e_q^{m\ln_q \eta}$.
From (\ref{generalized-meanvalue-product}),
\begin{eqnarray}
 \label{generalized-meanvalue-product2}
 \langle \widehat{B}_{t_1}\cdots\widehat{B}_{t_m}\rangle_q
 =
 K
(e_q^{\lambda_q  m})^{-1}
 \propto
 \frac{1}{e_q^{\lambda_q  m}},
\end{eqnarray}
where
$K=\frac{\eta}{N}$
is a number that remains fixed
in the classical (and thermodynamical) double limit
$\eta \rightarrow \infty$,
$N\rightarrow\infty$,
$\lambda_q=h_{\textrm{\scriptsize KS}}^{(q)}(T_{\kappa})
 = \ln_q \eta$
is a generalised Lyapunov exponent,
and
$\sum_{j_1,\ldots,j_m=1}^{M}
 \widetilde{\mu}(T_{t_1} A_{j_1}\cap\ldots \cap T_{t_m} A_{j_m})
 =
 \widetilde{\mu}(\vee_{i=1}^m T_{t_i}Q_{\scriptsize{\textrm{max}}})=1$.
Equations\ (\ref{uncorrelation deformed entropy 2})
and (\ref{generalized-meanvalue-product2})
describe the same $q$-exponential decay as a function of the time step,
labelled by $n$ and $m$ respectively.
In other words, in the classical limit
the cross-terms of the Weyl-Wigner expansions are neglected,
thus the $m$th point correlation function
$\langle \widehat{B}_{t_1}\cdots\widehat{B}_{t_m}\rangle_q$
and the asymptotical deformed uncorrelation are essentially the same quantity.
The approximation of a uniformly distributed perturbation can represent
a uniform environment that forces the system to decohere, for instance
the pendulum immersed in a continuum oscillator bath
(see p.\ 285 of \cite{Omn94}).

Now we are able to study some regimes of the fidelity decay
in a unified way.
%
\subsubsection{Lyapunov regime.}
$q\equiv q_{\textrm{\scriptsize fid}}=1$:
When the perturbation strength $\delta$ is sufficiently large,
(i.e., greater than the spacing of the energy levels)
then the Lyapunov regime takes place,
where the Loschmidt echo exponentially decays with a characteristic time
given by the Lyapunov exponent of the classical chaotic dynamics \cite{Jal01},
and
$
 \langle \widehat{B}_{t_1}\cdots\widehat{B}_{t_m}\rangle_q
 \propto
 e^{-\lambda  m},
$
$\lambda$ is the usual Lyapunov exponent.

\subsubsection{Regular regime.}
$q\equiv q_{\scriptsize \textrm{fid}}\neq1$:
In classically quasi-integrable systems the correlations between subsets
in phase space do not decay so fast as in the Lyapunov regime
but obey a power law $t^{-\frac{3D}{2}}$
($t^{-D}$ in the classical case), with $2D$ the dimension of phase space
\cite{Jac03}.
This derives from the diagonal part of $M(t)$
mainly determined by the decay of
$\langle \widehat{B}_{t_1}\cdots\widehat{B}_{t_m}\rangle$
(see equation\ (\ref{fidelity-series})).
Invoking the power law dependence of the correlations
(\ref{generalized-meanvalue-product2})
in the classical limit $\eta  \gg1$,
then $\lambda_q \ln_q \eta \gg 1$ and
the asymptotic behaviour of the $q$-exponential assures that
$e_q^{\lambda_q  m} \approx m^{\frac{1}{1-q}} \approx t^{-\frac{3D}{2}}$.
Thus, the entropic index for the regular regime is identified as
\begin{eqnarray}
 \label{regular-index}
 q_{\textrm{\scriptsize fid}}=1-\frac{2}{3D}.
\end{eqnarray}
The exponential decay is recovered with the increase
of the number of degrees of freedom $D$, in accordance with
a fast cancellation of the correlations in the macroscopic limit
$D\rightarrow\infty$.
The pendulum in a continuum oscillator bath, for instance,
displays an exponential decay of the correlations
in the macroscopic limit of an infinite number $N$ of
oscillators (see p.\ 287 of \cite{Omn94}).

\subsection{\label{subsec:application}
An example: the kicked rotator with absorption}

The kicked rotation with absorbing boundary conditions
is used as an instance showing that the generalised time scale $\tau_q$
is able to characterise the relaxation regime \cite{Bor91,Cas97}.
Differently from the ordinary kicked rotator,
the evolution operator of the present example is not unitary,
and thus their eigenvalues
are distributed inside the unit circle of the complex plane.
The model is described by the quantum map
\begin{eqnarray}\label{app1-map}
|\overline{\psi}\rangle=\widehat{U}|\psi\rangle=\widehat{P}
e^{-\frac{iT\widehat{n}^2}{4}}e^{-i\lambda\cos\widehat{\theta}}
e^{\frac{iT\widehat{n}^2}{4}}|\psi\rangle
\end{eqnarray}
where
$|\psi\rangle$ is an arbitrary state,
$|\overline{\psi}\rangle$ is the state after a time-step,
$\widehat{U}$ is the evolution operator,
$\widehat{P}$ is a projection operator over quantum state $n$ in the interval
$(-N/2,N/2)$ (being $N$ the size of the system),
$T$ is the period between two successive kicks,
$\lambda$ is the coupling strength of the kicks,
and $\theta$ is the angle of the pendulum with respect to the vertical.
The conjugated operators
$\widehat{n}$ and $\widehat{\theta}$ satisfy $[\widehat{n},\widehat{\theta}]=-i$
and the classical limit is characterised by the double limit
$\lambda \rightarrow \infty, T \rightarrow 0$
remaining constant the product $\lambda T$.

For studying quantum relaxation, Casati et al.\ \cite{Cas97} fixed
the classical chaos parameter $\lambda T = 7$ and the ratio $N/\lambda = 4$,
and the initial condition was chosen such that
only the level $n=0$ is populated.
In this case, the first $N$ complex eigenvalues of $\widehat{U}$ remain distributed
in a narrow ring of area $E_c$, which corresponds to the typical situation
in the scattering approach of open quantum systems.
The distance between the eigenvalues is $\delta\approx\sqrt{E_c/N}$,
where $\delta$ goes to zero and the density of poles becomes
a continuum in the classical limit.
For finite $N$, the separation of the poles can be resolved
after a relaxation time $\mathcal{T} \sim 1/\delta$, which implies
$\mathcal{T}\sim \sqrt{N}$.
In this regime, the cells of the quantum phase space
have a relative size $1/N$.

These results may be interpreted
by means of the graininess approach of Section \ref{subsec:graininess},
with
$1/N=\mu(A_i)=\frac{1}{\eta}$ for each rigid box $A_i$ ($i=1,\ldots,M$)
belonging to the maximal partition $Q_{\scriptsize{\textrm{max}}}$,
so $N=\eta$, which resembles the graininess of the quantum phase space,
$M=\eta$.
It follows that
$\mathcal{T}\sim \eta^{1/2}$,
i.e.,
$\mathcal{T}\approx \tau_{q=1/2}$
(with $\alpha=1/2$, see equation\ (\ref{q--time scale Heisenberg})),
a power law behaviour for the relaxation time.

\section{\label{sec:time-domain}
Towards a unified time domain scenario}

\begin{figure}[htb]
 \begin{center}
 \includegraphics[width=0.6\textwidth,clip=]{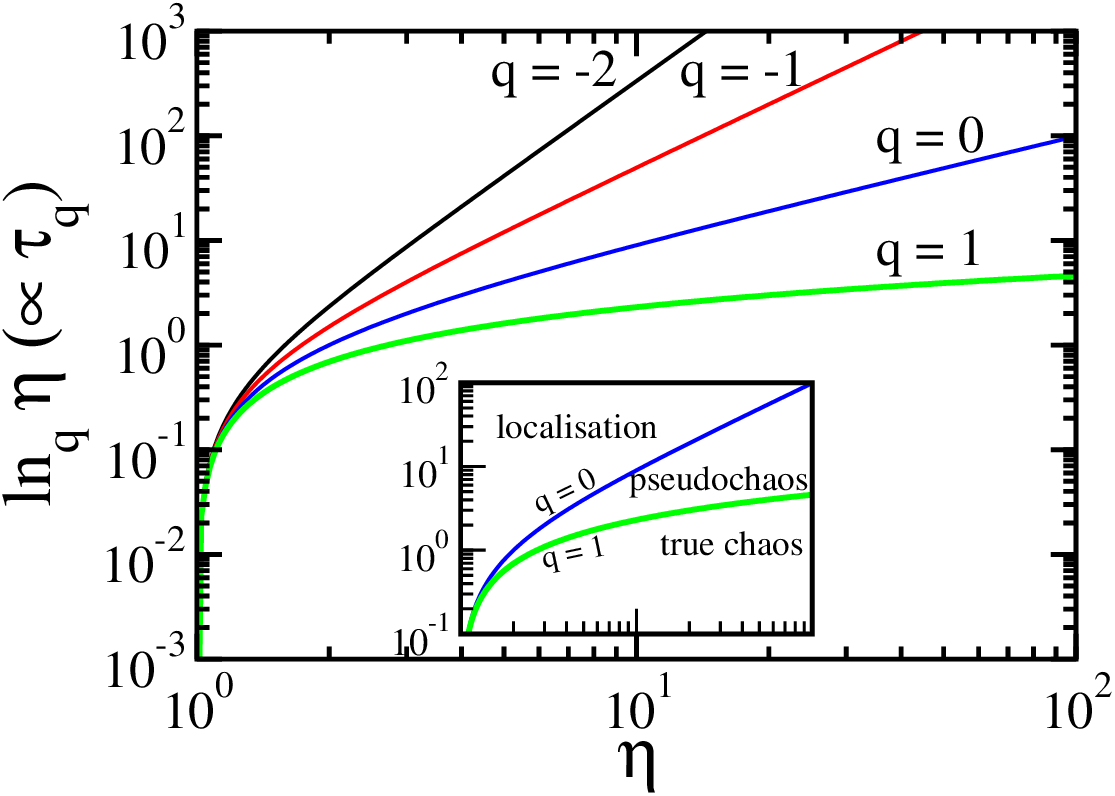}
 \end{center}
 \caption{\label{fig:qescalas}
  (Color online.)
  Log-log plot of $\ln_q \eta$ vs.\ $\eta$,
  with $\tau_q\propto \ln_q\eta$ and $\eta$
  is the quasiclassical parameter.
  The generalised time scale
  asymptotically resembles the general structure
  of the  quantum chaos dynamics
  (see Fig.\ 5 of \protect\cite{Cas95})
  in the classical limit $\eta\rightarrow\infty$,
  where the entropic index $q$ marks out the dynamics.
  Fig.\ 5 of \protect\cite{Cas95} plots
  the logarithm of the variables, while here
  we use logarithmic scales for axis---obviously
  both representations are equivalent.
  Inset: The same as the main frame,
  with the identification of three distinct regimes.
  Green solid line: logarithmic case $\tau_{q=1}$.
  Blue solid line: algebraic case, exemplified with $q=0$,
  that corresponds to the linear time scale
  (see equation\ (\protect\ref{q--time scale Heisenberg})).
  Any curve with $\tau_{q<1}$ produces
  the three region picture.
 }
\end{figure}

The concept of characteristic time scales in quantum dynamics
solves the ambiguity between the noncommutative double limit
(involving $t \rightarrow \infty$ and $\eta \rightarrow \infty$).
Numerical results show that every time scale must be obtained
for each regime separately.
In Ref. \cite{Cas95} it is asserted that the main peculiarity
of quantum chaos is the restriction to a finite interval,
called \emph{pseudochaos}, distinguished from the true chaos that
takes place in the classical limit
$\eta \rightarrow \infty$.
In virtue of this, it is concluded that the general structure of
quantum chaos dynamics can be expressed by the time scale curves on the plane
$(\eta,t)$ of Fig.\ 5 of \cite{Cas95}%
\footnote{
 N.B.: our $\eta$ and $\tau_q$ correspond to the symbols $q$
 and $t$ of Ref.\ \cite{Cas95} respectively. Since $q$ is generally referred
 as the entropic index in nonextensive statistics literature,
 we decided to follow this convention.%
 }.
Three well distinguished regions can be seen on their diagram:
(1) the power law region associated to the localisation phenomena,
(2) the region below the logarithm curve corresponding to true chaos regime,
and
(3) the intermediate region between (1) and (2) belonging to pseudochaos.
Interestingly, these regions are obtained by varying the entropic index $q$
in the formula of the generalised time scale $\tau_q$.
More precisely, for a given value of $q<1$, the power law region
is situated above (and on) the curve $\ln_q \eta$,
the true chaos zone is below (and on) $\ln_1 \eta$
(the usual logarithm),
and the pseudochaos region is placed in-between.
Fig.\ \ref{fig:qescalas} illustrates this characterisation
for some values of $q$.
Thus, the general structure of quantum chaos dynamics of
Fig.\ 5 of \cite{Cas95} can be reproduced by Fig.\ \ref{fig:qescalas}
in a unified way.
This unified scenario is identically reproduced in the classical limit
$\eta \gg 1$.
In fact, Casati and Chirikov asserted that the asymptotic limit
$t \rightarrow \infty$ must be taken conditionally in such a way that the ratio
$t/\tau_R(\eta)$, or $t/\tau_r(\eta)$, is fixed, depending on the dynamics.
This is precisely what we obtain using the generalised time scale:
the ratio $t/\tau_q(\eta)$ is kept fixed,
and the regime, either regular or chaotic, is defined by the entropic index $q$.

%
Now we provide a discussion about the connection between some approaches
in quantum chaos time scales and the framework presented in this paper.
Several previous works based on numerical experiments with simple models
as kicked rotators \cite{Haa01},
semiclassical propagation of wave packets \cite{Schu12},
randomness parameter in linear and nonlinear dynamical chaos \cite{Chi97},
and orderer quantisation \cite{Ang03},
among others, show that a framework for unifying the dynamical aspects
of quantum chaos would seem to be difficult to be defined in a single way.
Below we address some specific remarks reported on the literature,
and discuss them from the point of view of the present work.

\begin{itemize}
\item[$(1)$]
\emph{Universal nature of the time scales.}
The time scale diverges logarithmically with $h$
for classically chaotic flows
and it is algebraic in $h$ for the classical regular flow
(see \cite{Ang03} and references therein).
The present approach proposes the time scale to be expressed by
a $q$-logarithm function, equation\ (\ref{q--time scale})
(the entropic index could be more properly renamed within the present context
as $q_{\textrm{\tiny QC}}$, {\scriptsize QC} stands for quantum chaos),
that interpolates between the logarithmic regime
(with $q_{\textrm{\tiny QC}} = 1$)
and the algebraic one
(with $q_{\textrm{\tiny QC}} = 1-\alpha$),
in the classical limit $\eta \rightarrow \infty$.

\item[$(2)$]
\emph{Kolmogorov-Sinai time.}
The exponential increase of any small initial volume $\Delta \Gamma_0$
within the phase space,
$\Delta \Gamma(t)=\Delta \Gamma_0 e^{h_{\textrm{\tiny KS}}t}$,
implies that the whole phase space is filled ($\Delta \Gamma \approx 1$)
after a time of the order
$t_{0} = \frac{1}{h_{\textrm{\tiny KS}}} \log \frac{1}{\Delta \Gamma_0}$,
and thus the relaxation time might be proportional to
$\frac{1}{h_{\textrm{\tiny KS}}}$
(see \cite{Del97} and references therein).
The generalised time scale here introduced points towards
a $q$-exponential spreading of a small phase volume
$\Delta \Gamma_0$ over a region
$\Delta \Gamma(t) = \Delta \Gamma_0 e_q^{h_{\textrm{\tiny KS}}^{(q)}(T)t}$
after a time $t$.
It follows straightforwardly that
\begin{eqnarray}
 \label{q KS time}
 \tau_{\textrm{\tiny KS}}^{(q)}(T) = \frac{1}{h_{\textrm{\tiny KS}}^{(q)}(T)}
\end{eqnarray}
constitutes a generalisation of the Kolmogorov-Sinai time
$\tau_{\textrm{\tiny KS}}(T) = \frac{1}{h_{\textrm{\tiny KS}}(T)}$
for processes satisfying the asymptotical deformed uncorrelation
(\ref{uncorrelation deformed entropy}).

\item[$(3)$]
\emph{Modifications of the quantum chaos theory.}
There are claims that
the quantum chaos phenomenon
seems to ask for a possibly difficult mathematical modification of the theory
for a finite time (see \cite{Cas95}, Sec.\ 3.3, p.\ 18).
The main framework of the theory is preserved
by using the generalised time scale $\tau_q$,
and
the generalised KS-entropy
(\ref{deformed KS entropy}).
The kind of correlation, if any, between the subsets of the quantum phase space
defines which time scale is to be used.
The particular case of the asymptotical deformed uncorrelation,
equation\ (\ref{uncorrelation deformed entropy}),
leads to the present $\tau_q$ generalisation.
\end{itemize}

\section{\label{sec:conclusions}Conclusions}

We have presented a redefinition of the quantum chaos time scales
by means of a generalised time scale, $\tau_q$
(henceforth we use $q_{\textrm{\tiny{QC}}} \equiv q$),
motivated by the nonextensive statistics,
which contains the relaxation and random regimes as particular cases.
We have obtained
the generalised time scale by using four ingredients:
(i) a generalised KS-entropy (equation\ (\ref{deformed KS entropy})),
(ii) the rescaled time property (equation\ (\ref{rescaled KS entropy})),
(iii) the asymptotical deformed uncorrelation
      (equation\ (\ref{uncorrelation deformed entropy})), and
(iv) the graininess of the quantum phase space
     (equation\ (\ref{graininess})).

Lyapunov and regular regimes for the fidelity decay has been obtained
as a consequence of the generalisation of the $m$th point correlation function
through the $q$-product,
equation\ (\ref{generalized-meanvalue-product}),
for a uniformly distributed perturbation in the classical limit.

The deformed uncorrelation introduced by
equation\ (\ref{uncorrelation deformed entropy})
is not claimed as a necessary, but a sufficient condition
to obtain the generalised time scale.
A physical justification of it is still absent,
however, it is intriguing that this
hypothesis
leads to the unified
scenario here presented.
Other correlations different from that given by
equation\ (\ref{uncorrelation deformed entropy})
will possibly lead to different generalisations for the time scales.

Within the framework of nonextensive statistical mechanics
(see Section 3.3.4 of \cite{tsallis-book}),
the entropic index $q_{\textrm{\scriptsize{ent}}}$
of the generalised entropy $S_{q_{\textrm{\scriptsize{ent}}}}$
is ultimately defined by the correlations
between the subsystems of a composite system,
that stem from its dynamics.
For strongly chaotic systems,
i.e., systems with exponential sensitivity to the initial conditions
(with at least one positive Lyapunov exponent),
corresponding to independent subsystems
(i.e., uncorrelated subsystems, but it also asymptotically works
 for weakly correlated subsystems),
the effective volume $W$ of the classical phase space
exponentially increases with the size $N$ of the system
(we mean, a system composed by $N$ equal subsystems).
The Boltzmann-Gibbs entropy (for the equiprobable case)
$S_{q_{\textrm{\scriptsize{ent}}}=1} =S_{\textrm{\scriptsize{BG}}} = k_B \ln W$
is thus extensive,
i.e., $S_{\textrm{\scriptsize{BG}}}(N) = N S_{\textrm{\scriptsize{BG}}}(1)$.
For weakly chaotic systems,
i.e., systems with a power law sensitivity to the initial conditions
(with vanishing maximal Lyapunov exponent),
for which
the effective volume $W$ of the classical phase space
increases according to a power law of the size $N$ of the system
(say, $W \sim N^d$, with $d>0$),
the extensivity of the entropy
$S_{q_{\textrm{\scriptsize{ent}}}<1}(N) \propto
 N S_{q_{\textrm{\scriptsize{ent}}}<1}(1)$
can only be attained for an entropy with $q_{\textrm{\scriptsize{ent}}} = 1-1/d < 1$
(the Boltzmann-Gibbs entropy
$S_{\textrm{\scriptsize{BG}}}(N) \propto \ln N$,
thus it is not extensive in this case).
%
Note that both cases, the strongly and the weakly chaotic systems,
can be unified through a $q$-exponential increase of $W$ with $N$.

This structure is impressively similar to the present generalisation
of the time scales.
For the chaotic regime, the quasiclassical parameter $\eta=\frac{I}{h^D}$,
--- that plays the role of the normalised volume $W$
(the number of states) of the classical phase space ---,
exponentially increases with the time scale
$\tau_{q_{\textrm{\tiny{QC}}}=1}=\tau_r$
(the inverse of equation\ (\ref{logarithmic time scale})).
For the regular regime, $\eta$ increases according to a power law
of the time scale $\tau_{q_{\textrm{\tiny{QC}}}<1}=\tau_R$
(the inverse of equation\ \ref{power law time scale}).
In this paralleling, the generalised time scale
$\tau_{q_{\textrm{\tiny{QC}}}}$
plays an analogous role to the size $N$ of the system.
Similarly to the index $q_{\textrm{\scriptsize{ent}}}$,
the index $q_{\textrm{\tiny{QC}}}$
is also determined by the (quantum) dynamics,
and it is also less than 1.
The connection between
the asymptotic scale invariance correlations
and
the asymptotic scale-free occupation of the phase space,
remarked by \cite{tsallis-gell-mann-sato--2005}
in a classical formulation,
also seems to be present in the semiclassical approach.
The nonextensive statistical mechanics,
that unifies aspects of the classical chaos dynamics,
may also unify the quantum chaos dynamics.

\section*{Acknowledgements}
E.P.B. acknowledges the support of the National Institute of Science and
Technology for Complex Systems (INCT-SC).
I.S.G.  acknowledges fellowships received from  CONICET
(at Universidad Nacional de La Plata)
and from CAPES / INCT-SC (at Universidade Federal da Bahia).

\end{document}